\documentclass[]{aa}
\usepackage{txfonts}
\newcommand{\oversim}[2]{\protect{\mbox{\lower0.5ex\vbox{%
   \baselineskip=0pt\lineskip=0.2ex
   \ialign{$\mathsurround=0pt #1\hfil##\hfil$\crcr#2\crcr\sim\crcr}}}}} 
\newcommand{\simless} {\mbox{$\,\mathrel{\mathpalette\oversim<}\,$}} 
\begin{document}

\title{Limits on the primordial stellar multiplicity}

\author{Simon\,P.\,Goodwin$^1$ \and Pavel\,Kroupa$^{2,3}$}

\authorrunning{S.\,P.\,Goodwin \and P.\,Kroupa}

\offprints{Simon.Goodwin@astro.cf.ac.uk}

\institute{$^1$School of Physics \& Astronomy, Cardiff University, 
5 The Parade, Cardiff, CF24 3YB, UK \\ $^2$ Sternwarte,
Universit\"at Bonn, Auf dem H\"ugel 71, D~53121 Bonn, Germany\\
$^3$ Rhine Stellar Dynamical Network}

\date{}

\abstract{ Most stars - especially young stars - are observed to be in
  multiple systems.  Dynamical evolution is unable to pair stars
  efficiently, which leads to the conclusion that star-forming cores
  must usually fragment into $\geq 2$ stars.  However, the dynamical
  decay of systems with $\geq 3$ or $4$ stars would result in a large 
  single-star population that is not seen in the young stellar
  population.  Additionally, ejections would produce a significant 
  population of close binaries that are not observed.  This leads to a 
  strong constraint on star formation theories that cores must typically
  produce only 2 or 3~stars. This conclusion is in sharp disagreement
  with the results of currently available numerical simulations that
  follow the fragmentation of molecular cores and typically predict
  the formation of 5--10 seeds per core.  In addition, 
  open cluster remnants may account for the majority of observed 
  highly hierarchical higher-order multiple systems in the field.
  
   }

\maketitle

\section{Introduction}

In the Galactic field most solar-type stars are in multiple systems
with a multiplicity fraction, $f_{\rm mult} = (B+T+Q)/(S+B+T+Q) = 
0.58$ where $S, B, T, Q$ are the number of single, binary, triple 
and quadruple systems, respectively (Duquennoy \& Mayor 1991,
including significant corrections for incompleteness from the raw
value of 0.49).  Recent studies have found that
$f_{\rm mult}$ in the field may be somewhat higher than this (Quist 
\& Lindegren 2000; S\"oderhjelm 2000).  Duquennoy \& Mayor (1991) find 
the uncorrected ratio of systems to be $S:B:T:Q=1.28:1:0.175:0.05$ (see also  
Tokovinin \& Smekhov 2002).

There is evidence that $f_{\rm mult}$ is significantly higher among 
young stars than in the field (e.g. Leinert et al.  1993; Mathieu 
1994; Duch{\^ e}ne 1999; Bodenheimer et al. 2000; Patience et al. 
2002; Kroupa et al. 2003; Duch{\^ e}ne et al. 2004; Haisch et al. 
2004).  In Taurus essentially all stars between 0.3 
and 1 $M_\odot$ are in multiple systems with separations extending 
to about 1500~au (Leinert et al. 1993; Duch\^ene et al. 2004; Haisch 
et al. 2004).  Samples of T~Tauri stars in other nearby star-forming 
regions also show that T~Tauri stars generally have a high 
$f_{\rm mult}$ and a bias towards larger separations than the 
main-sequence sample (Duch\^ene 1999; Patience et al. 2002; Duch\^ene 
et al. 2004; Haisch et al. 2004).  Such samples are generally biased 
towards large separations ($\sim 100$ to $1000 - 3000$ au) and may 
miss many close binaries.  Generally it is found that $f_{\rm mult}$
in the sampled separation ranges is approximately twice that in the 
Galactic field and if extrapolated to the whole separation range 
implies $f_{\rm mult} \sim 1$.

It has been shown that it is not possible to reproduce the observed
$f_{\rm mult}$ through the dynamical evolution of star clusters that are
born with a single-star population (Kroupa 1995a).  Dynamical
interactions are able to disrupt many wide binaries, but are not able 
to pair stars efficiently or significantly change the properties of 
close binaries (e.g. Kroupa 1995a,b; Kroupa et al. 1999; Kroupa \& 
Burkert 2001; Kroupa et al. 2001).  This leads us to the conclusion 
that the majority of stars must form in multiple systems.

All stars form in dense molecular cores (Andr\'e et al. 2000), and
most of these star-forming cores are found within larger complexes
that will become star clusters.  It is these star-forming cores that
must fragment into multiple objects.  Recent simulations have shown
that cores are able to fragment if there is a small amount of initial
rotation (eg. Burkert \& Bodenheimer 1996), with compression due to an
external pressure (eg. Hennebelle et al.  2003, 2004), or if they
contain high- (eg. Bate et al. 2002, 2003; Delgado Donate et al. 2004)
or low- (eg. Goodwin et al. 2004a,b) levels of turbulence. As reviewed
by Duch{\^ e}ne et al. (2004) the models predict the cores to
generally fragment into $N=5-10$ pre-stellar objects.  But Duch{\^
e}ne et al. point out that this appears to contradict the
observational findings according to which no, or very little,
higher-order sub-clustering within cores is evident.

In this paper we re-address this problem by considering the elementary
constraints on the number of stars that can form within a core posed
by the decay of higher-order multiple systems.  Limits on the
number of objects that usually form within cores provide very strong
constraints on theories of star formation.

\section{Constraints on the fragmentation of cores}

As a core collapses it is initially isothermal as it is optically thin
to its own radiation.  At a critical density of $\rho_{\rm crit}
\sim 10^{-13}$ g cm$^{-3}$ for a $1\,M_\odot$ core, the gas
becomes optically thick and enters an adiabatic regime (eg. Larson
1969; Tohline 1982; Masunaga \& Inutsuka 2000).  This results in a
minimum Jeans mass being reached around $\rho_{\rm crit}$ of $M_{\rm
min} \sim 10^{-2} M_\odot$.  It is at $\sim \rho_{\rm crit}$ that
fragmentation is expected to occur producing fragments of mass $M_{\rm
min}$.  Thus the length scale of the initial multiple systems is
expected to be

\begin{equation}
R_{\rm form} < \left( \frac{3M_{\rm core}}{4\pi \rho_{\rm crit}}
\right)^{1/3} < 125 (M_{\rm core}/M_\odot)^{1/3} \,\,\, {\rm au},
\end{equation}

\noindent where $M_{\rm core}$ is the mass of the pre-stellar core.
This scenario is supported by the observed peak in the T Tauri binary
distribution at $\sim 100$ au (eg. Mathieu 1994; Patience et
al. 2002). 

\subsection{Dynamical decay}

Any (non-hierarchical) multiple system with $N \geq 3$ members is 
generally unstable to decay with a half-life of $t_{\rm decay} 
\sim 80$ crossing times,

\begin{equation}
t_{\rm decay} \sim 14 \left( \frac{R}{{\rm au}} \right)^{3/2} \left(
\frac{M_{\rm stars}}{M_\odot} \right)^{-1/2} \,\,\,\,\, {\rm yrs},
\label{eq:tdec}
\end{equation}

\noindent where $R$ is the size of the system and $M_{\rm stars}$ the
mass of the components (Anosova 1986; see also Reipurth \& Clarke
2001).  A typical decay time-scale for a system with $R=250$~au
($2R_{\rm form}$) and $M_{\rm stars}=1 M_\odot$ is then of order 
55~kyr.  The decay of multiple systems has two important consequences: 
the preferential ejection of the lowest-mass member of the system 
(Anosova 1986; Sterzik \& Durisen 2003), and the reduction of the 
semi-major axis of the remaining binary by a factor of 2-4 
(Anosova 1986).  The reduction in the semi-major axis is significantly 
greater if ejections occur whilst accretion is on-going (Umbreit et
al. 2005).  Thus, for decaying higher-order systems $t_{\rm decay}$
falls rapidly as each ejection reduces the size of the system
significantly (as $t_{\rm decay} \propto R^{3/2}$).  {\it The
complete decay of a higher-order system to one binary and a number of 
single stars is expected to occur in less than~0.1~Myr}. 

This raises the possibility that at least some classical and
weak-lined T~Tauri sources have been recently ejected from embedded
multiple systems and are coeval with those systems, not older than
them.  Ejected stars would leave a core in $\sim 100$~kyr and
depending on the size of a disc that they are able to retain during
ejection would appear indistinguishable from classical and
weak-lined T~Tauri sources.  {\it Thus the classical picture that 
protostellar sources evolve from class 0 $\rightarrow$ I 
$\rightarrow$ II $\rightarrow$ III may need revision} as it does not 
include (a) the possibility of creating later-type sources immediately 
through ejections, and (b) the lack of a one-to-one relationship 
between the numbers of cores and the numbers of later-type sources.
In particular, the relative lifetime estimates of young stars assume a
one-to-one correlation between embedded and non-embedded sources, the
differences in relative numbers being due to the different durations
of each phase (e.g. Greene et al. 1994) which may be incorrect.

\subsection{Diluting the multiplicity fraction}

Ejections are predominantly of single stars (see Bate et al.  2002,
2003; Delgado Donate et al. 2004; Goodwin et al. 2004a,b for 
examples), and therefore ejections dilute $f_{\rm mult}$ 
(e.g. Bouvier et al. 2001).  A cluster which is (initially) composed of
triple systems would have a $f_{\rm mult}$ of unity, however after the
decay of these systems 1/3 of stars will be single, and 2/3 in
binaries reducing $f_{\rm mult}$ to 0.5. Similarly, a population of
quadruples will decay within less than $10^5$~yr to a population with
$f_{\rm mult} = 1/3$, assuming each quadruple decays to two single
stars and one binary.  

The assumption is made here that the distances
between members of the multiple system are approximately equal and
that all of the members form at the same time.  Simulations of
multiple system formation within cores show that often hierarchical
systems can form with the members appearing at slightly different
times (see Delgado Donate et al. 2004; Goodwin et al. 2004a,b). These
multiples remain stable for at least 10~Myr (Delgado Donate et
al. 2004).  However, higher-order multiples in these simulations
generally form as the result of the decay of even higher-order systems
($N \geq 5$ stars), so that $f_{\rm mult}$ is low for the whole
population.

In contrast to the above expectations from dynamical decay,
star-forming regions are typically observed to have an extremely 
high $f_{\rm mult}$ (see Section 1).  In addition to finding a
significantly high $f_{\rm mult}$, Duch{\^e}ne et al. (2004) 
find no significant difference in $f_{\rm mult}$ and the separation
distribution between the physically different
star-forming regions $\rho$~Oph and Taurus, but a marked difference to
the Galactic-field sample.  Duch\^ene et al. (2004; also see Haisch et
al. 2004) find that $f_{\rm mult}$ is significantly higher in young
embedded sources ($38 \pm 8$\%) than in older flat-spectrum sources
($22 \pm 3$\%) in both Taurus and $\rho$~Oph (in both cases this is
significantly higher than the field $f_{\rm mult}$ of 14\% for the
same separation range). This suggests that there has been some
dynamical decay of young higher-order multiple systems on a time-scale
$<1$~Myr.  This observation also suggests that some of the late-type
sources are single stars ejected from early-type multiple systems.

In a cluster such as the Orion Nebula Cluster which has a $f_{\rm
mult} \sim 0.6$ very similar to the field (Duch{\^e}ne 1999), about
one star could have been ejected per binary system, i.e. cores may
typically produce unstable triple systems, or alternatively, that
roughly $1/3$ cores only produced a single star while $2/3$ produced
binaries.  This leads to a basic constraint on star formation within
cores in Orion of $1 \leq N \leq 3$.  On the other hand, without the
need to invoke a different population of cores that form different
numbers of objects to other star-forming regions, many (wide)
primordial binaries can be destroyed by dynamical interactions in 
such a cluster (Kroupa et al. 2001).  It is thus possible to match 
the Orion and the Galactic field $f_{\rm mult}$ and separation 
distribution if the initial population is comprised mainly of 
binary systems with the T~Tauri separation distribution (Kroupa 1995a,b).

It may be that ejected stars have escaped from the cluster (or at
least travelled far enough not to be included in samples) and so
observations are of preferentially high $f_{\rm mult}$ regions.
However, the ejection velocity of stars is $\sim 1$ km s$^{-1}$ (Bate
et al. 2002,2003; Delgado Donate et al. 2004; Goodwin et al. 2004a)
which is comparable with the velocity dispersion of star-forming
cores. Hence star-forming regions would be expected to be mixed.
Also, the potential well of larger clusters (such as Orion) is too
deep to allow the loss of stars ejected at such velocities.  We
conclude that ejected stars and their progenitor systems should
normally be coincident.

The generally high $f_{\rm mult}$ for pre-main sequence late-type 
stars uncovers {\it an elementary discrepancy if cloud cores produce 
$N>3$~stars}.

\subsection{Primordial triples}

A non-negligible fraction of T~Tauri stars in the surveyed regions do
appear to be hierarchical triples, probably $>25\%$.  Leinert et al.
(1993) find 4 higher-order multiples from 45 systems in Taurus, at
least 7 more are known with possibly more currently undetected.  More
recently, Koresko (2002) has used speckle holography at $2\mu$m at
Keck to search for tertiary companions among 14 pre-main sequence
visual binary systems finding that possibly up to half of the imaged
objects may be triples, although clear identification of tertiary
companions was possible only in two of the possible~7 cases.
Noticeably, a large fraction of Koresko's sample lies in the Ophiuchus
cloud, a region that may be seen as a moderately rich cluster, clearly
different from the Taurus cloud surveyed by Leinert et al., suggesting
that this is a relatively general feature and not a ``single cloud''
phenomenon. In yet more dispersed star-forming regions (the TW Hya and
MBM 12 associations), Brandeker et al. (2003) found a similarly high
ratio of triples-to-binaries, though with limited statistical
significance.

In the Galactic field, Tokovinin \& Smekhov (2002) suggest that
20-30\% of multiple systems are higher-order hierarchical systems,
consistent with the fraction of T~Tauri stars suggesting that many of
these young systems are stable as found in the simulations of
Delgado Donate et al. (2004). The fraction of triple and quadruple 
systems per star, ${\cal F}=(T+Q)/(S+2B+3T+4Q) = 0.05$ for the 
Galactic field, this corresponds to a higher-order multiplicity
fraction of $(T+Q)/(B+T+Q)=18\%$ of all multiple systems.

The very high frequency of primordial binary and triple systems in 
the Taurus and $\rho$~Oph star-forming regions suggest that very few 
ejections from multiple systems have occurred.  This implies 
that {\it stars must form in groups of 2 or 3 within cores.}

\subsection{Hardening and close binaries}

Further support for our conclusion that $2 \leq N \leq3$ is provided
by the separation distribution of systems.  There are many binaries in
Taurus with wide separations $> 100$ au (roughly double the
number found in the field) which also implies that few
ejections can have occurred, as ejections rapidly harden binaries so
that their semi-major axes are $<10$ au -- from an initial separation
of 100~au, 2~or 3~ejections are sufficient (Anosova 1986; Umbreit et
al. 2005).

Ejections are expected to occur during the main (class 0) accretion
phase from basic decay-time arguments (see above), and also as seen in star 
formation simulations (Bate et al. 2002,2003; Delgado Donate et al. 2004;
Goodwin et al. 2004a,b; see also Reipurth \& Clarke 2001).  Ejections
during this phase are far more effective at hardening binary systems
(see Umbreit et al. 2005). This occurs as ejections harden a
lower-mass system than the final system thus requiring less energy.
In addition, early hardening tends to lead to a population of 
equal-mass binaries, as the close low-mass component is more able to accrete 
high angular momentum material from the disc thus pushing the mass 
ratio towards unity (Whitworth et al. 1995; Bate \& Bonnell 1997).
Such systems are not observed in significant numbers in the 
field (Duquennoy \& Mayor 1991; Mazeh et al. 1992).

It is not expected that stars would form much closer together than
$\sim 20$ -- 50~au (the separation between two minimum Jeans mass
objects at $\sim \rho_{\rm crit}$, see Eqn. 1).  Ejections are thought
to be the primary method by which these systems are hardened.
However, these systems are unlikely to have been
hardened during the main accretion phase as they do not strongly favour
equal-mass binaries, although short-period binaries do favour more 
similar component masses than wider binaries (Mazeh et al. 1992).
They may however form from the (fairly late) decay of some fraction of
the triple systems.

Thus, in order to prevent a large population of close binaries
being formed from the decay of unstable triple systems, {\em $N=2$ must
be a more common outcome of star formation than $N=3$.}

\subsection{Fragmentation consistent with data}

We propose the following picture that is consistent with the data.  Of
every 100 star forming cores within a cloud roughly 60 will form 
binary and 40 will form triple systems.  Of those which form triples,
around 25 are long-lived hierarchical systems, whilst the other 15 
triples decay to 15 binaries and 15 single low-mass stars and/or brown 
dwarfs.  The 15 binary systems that result from dynamical decay form 
the close binary population.

This leads to 115 systems, 15 of which are single stars, 75 are
binaries and 25 triples with $f_{\rm mult}=100/115=0.87$ for the whole
population.  As the ejected stars were the lowest mass members of 
the triple systems, $f_{\rm mult}$ remains $\sim 1$ for larger $>0.3 -
0.5 M_\odot$ stars, but is low among low-mass stars and brown dwarfs.
Such a primordial population would then either remain essentially unevolved 
in Taurus-like star-forming regions, or would evolve dynamically in 
dense Orion-like environments on a cluster crossing-time scale (Kroupa 2000; 
Kroupa et al. 2001; Kroupa et al. 2003).

\subsection{Brown dwarfs}

The dynamical ejection of sub-stellar embryos has been suggested as a
formation mechanism for brown dwarfs (Reipurth 2000; Reipurth \& Clarke
2001).  Ejections are expected to occur during the main accretion
phase (see above), indeed brown dwarfs must be ejected shortly after
they form as the accretion rates of very young sources rapidly grow 
the embryos to stellar masses (which has been suggested as an
explanation of the brown dwarf desert, Reipurth \& Clarke 2003).  The 
ejection of brown dwarfs, despite their low mass, would still have a
significant effect on the separation (and hence mass ratios) of 
young multiple systems if they occur when all of the components have 
not accreted a significant amount of mass (see also Umbreit et
al. 2005).  Thus, ejections of brown dwarfs during the main 
accretion and star formation phase\footnote{It is expected that 
star formation would become difficult after the class 0 phase as most 
of the gas available for star formation has already 
been accreted.} must be limited in order to avoid creating a large 
population of close, equal-mass binaries. 

The available star-count data in the Taurus star-forming region 
and in Orion constrains this process to about one brown dwarf being 
ejected per four stars in both environments (Kroupa \& Bouvier
2003). So this scenario is not ruled-out as a brown dwarf formation
mechanism, especially if embryo formation can be delayed and brown
dwarfs ejected from systems whose other members have a significantly
higher mass, in which case the hardening of the parent system would be
reduced.  It should be emphasised that, even if {\em all} brown 
dwarfs are ejected embryos, the limit of one brown dwarf per 4 stars
(or per two stellar binary systems) still constrains the total number
of ejections to be low.

\subsection{Cluster remnants as a source of high-order systems}

As stated above, many of the observed $N=3$ systems in the field may
be primordial, i.e. may have formed as stable hierarchical systems
from molecular cloud cores. However, some triple and perhaps most
higher-order multiple systems may also be formed by star cluster
remnants.

de La Fuente Marcos (1997, 1998) points out that cluster remnants are
characterised by a few binaries, ie. highly hierarchical high-order
multiples, and that the number of such objects can be large given that
the majority of Galactic-field stars form in modest clusters.
Evaporating clusters decay into highly hierarchical systems of $N
\simless 15$ members.  These remnants will presumably decay further
over time possibly producing quadruple or quintuple systems. We can
estimate the number of highly hierarchical $N>3$ systems under this
scenario as follows:

For a power-law embedded cluster mass function, $\xi_{\rm ecl} = k\,
M_{\rm ecl}^{-\beta}$, where $\xi_{\rm ecl}\,dM_{\rm ecl}$ is the
number of embedded clusters with mass in the interval $M_{\rm ecl},
M_{\rm ecl}+{\rm d}M_{\rm ecl}$ and $M_{\rm ecl}$ is the mass in
embedded cluster stars, the number of hierarchical high-order stellar
systems is equal to the number of embedded clusters in a mass interval
$M_{\rm ecl,min}$ to $M_{\rm ecl,max}$, $N_{\rm ecl}=\int_{M_{\rm
ecl,min}}^{M_{\rm ecl,max}}\xi_{\rm ecl}\,{\rm d}M_{\rm ecl}$, since
each cluster leaves one such system. The number of stars formed in the
embedded cluster population during some epoch is $N_{\rm
st}=\int_{M_{\rm ecl,min}}^{M_{\rm ecl,max}} (M_{\rm ecl}/m_{\rm
av})\,\xi_{\rm ecl}\,{\rm d}M_{\rm ecl}$, where $m_{\rm
av}=0.4\,M_\odot$ is the average stellar mass which we take to be
constant for the present estimate.  Thus, for $M_{\rm
ecl,min}=5\,M_\odot$ (a cluster of a dozen stars, Taurus-like object,
the least-massive star-forming ``unit'' known) and $M_{\rm
ecl,max}=500\,M_\odot$ (since most Galactic-field stars appear to stem
from rather modest clusters, Kroupa 1995a; Adams \& Myers 2001) and
for $\beta=2$ (Lada \& Lada 2003) we have $N_{\rm ecl}=0.19$ and
$N_{\rm st}=4.43$ so that the above defined quantity ${\cal F}=N_{\rm
ecl}/N_{\rm st}\sim 4$~\%.  {\it Therefore the number of
cluster remnants may be completely sufficient to account for all
observed higher-order multiple systems.}

\section{Conclusions}

The high multiplicity fraction $f_{\rm mult}$ of pre-main sequence
stars implies that most stars form in multiple systems within dense
molecular cores.  The decay of young multiple systems generally within 
$<0.1$~Myr has two main effects: diluting the multiple population
through the ejection of predominantly single stars, and hardening the
remaining systems.

The observed large fraction of $\sim 10^6$~yr old, wide binary and
hierarchical triple systems implies that (a) not much dilution of
the binary population can have occurred through ejections, and (b) few
ejections can have occurred as systems have not been significantly
hardened.  

This suggests that most cores form only $2 \leq N \leq 3$
stars. Typical numbers that are consistent with observational
constraints are that of 100 birth systems 40 are triple and 60
binaries.  Of the 40 triples 25 are long-lived hierarchies while 15
decay to 15 close binaries and 15 single stars. This is then
the primordial population as seen dynamically unevolved in Taurus and
$\rho$~Oph (Kroupa et al. 2003).  The destruction of binary systems by
dynamical interactions in dense clusters such as the Orion Nebula
cluster further dilutes the binary fraction (Kroupa et al. 2001).
Wide and low-mass binaries are preferentially destroyed, and the
majority of single low-mass stars can be explained through this
process and do not require ejections.  Such dynamical disruptions
occur in the early stages of cluster evolution and explain the rapid
drop in the binary fraction observed between T~Tauri stars 
and older clusters and the
Galactic-field main sequence population.  The observed
variations may be entirely understood through dynamical disruption of
wide binary systems in differently crowded regions.

This would imply a remarkable invariance of core fragmentation to
differing star-formation conditions. Motte \& Andre (2001) have shown
that protostellar cores show different sizes, densities and edge
sharpness if they are located in cluster-like environments or sparser
associations. Nevertheless, the remarkable success in describing a
variety of stellar populations with the same primordial binary
population does suggest that core fragmentation may indeed be about
invariant on the scale of a few hundred~au or less.  Such an
invariance may not be unexpected as, whilst large-scale core
properties vary significantly, it is only after collapse through
several orders of magnitude in density that stars form, and such a
collapse may make many large-scale properties irrelevant.

Ejections from cloud cores can lead to brown-dwarfs as ejected
embryos, but the frequency of such events is already limited by
available data to not more than 1~brown dwarf per two stellar
binaries (or one per four stars).  

Cluster remnants can form highly hierarchical high-order multiple
systems. These contribute to the observed number of $N\ge3$ stellar
systems. Cluster remnants may account for all observed higher-order
($N>3$) multiple systems.  Therefore, possibly only a small fraction
of molecular cloud cores have to be able to form long-lived stable
hierarchical higher-order multiple systems.

These conclusions place strong constraints on theories of star 
formation.  For any theory of star formation to match observations 
the majority of cores {\em must} fragment into multiple objects.
However, they can usually only fragment into 2 or~3 stars. The 
currently available theoretical results appear to be inconsistent 
with this, as the cloud-core fragmentation calculations typically 
form $N=5-10$ fragments per core.  This discrepancy may not be unexpected 
because the theoretical work misses the very essential physics of 
stellar feedback, and cannot take into account magnetic forces that 
may play a role in the fragmentation of a core, and simplifies the 
complex and important thermal physics (in particular simplifying 
the equation of state of the gas).  

\begin{acknowledgements}

SPG is supported by a UKAFF Fellowship.

\end{acknowledgements}

\end{document}